\documentclass[global]{svjour}
  
\usepackage{graphicx}
\usepackage{siunitx}
\usepackage{braket}
\usepackage{lineno}
\usepackage{ulem} 

\begin{document}

\title{JOKARUS---Design of a compact optical iodine frequency reference for a sounding rocket mission}

\author{Vladimir Schkolnik\inst{1}\thanks{$^{\dagger}$ contributed equally to this work} \and Klaus D\"oringshoff\inst{1}$^{\dagger}$ \and Franz Balthasar Gutsch\inst{1} \and Markus Oswald\inst{2} \and Thilo Schuldt\inst{3} \and Claus Braxmaier\inst{2,3} \and Matthias Lezius\inst{4} \and Ronald Holzwarth\inst{4} \and Christian K\"urbis\inst{5} \and Ahmad Bawamia\inst{5} \and Markus Krutzik\inst{1} \and Achim Peters\inst{1}  
}


\institute{ Institut f\"{u}r Physik, Humboldt-Universit\"{a}t zu Berlin, Newtonstr. 15, 12489 Berlin, Germany\\ \email{vladimir.schkolnik@physik.hu-berlin.de} 
\and Zentrum f\"{u}r angewandte Raumfahrttechnologie und Mikrogravitation (ZARM), Universit\"{a}t Bremen, Am Fallturm, 28359 Bremen, Germany
\and Deutsches Zentrum f\"ur Luft- und Raumfahrt (DLR), Institut f\"ur Raumfahrtsysteme, Linzer Straße 1, 28359 Bremen, Germany
\and Menlo Systems GmbH, Am Klopferspitz 19a, 82153 Martinsried, Germany
\and Ferdinand-Braun-Institut, Leibniz-Institut f\"{u}r H\"{o}chstfrequenztechnik,
Gustav-Kirchhoff-Str. 4, 12489 Berlin, Germany }

\date{}

\maketitle

\begin{abstract}
We present the design of a compact absolute optical frequency reference for space applications based on hyperfine transitions in molecular iodine  with a targeted fractional frequency instability of better than \num{3e-14}. It is based on a micro-integrated extended cavity diode laser with integrated optical amplifier, fiber pigtailed second harmonic generation wave-guide modules, and a quasi-monolithic spectroscopy setup with operating electronics. 
The instrument described here is scheduled for launch end of 2017 aboard the TEXUS\,54 sounding rocket as an important qualification step towards space application of iodine frequency references and related technologies. The payload will operate autonomously and its optical frequency will be compared to an optical frequency comb  during its space flight.

\end{abstract}



\section{Sounding rockets as steppingstone for space-borne laser systems}
\label{sec:rocket}

Frequency stable laser systems are a mandatory key technology for future space missions using optical and quantum-optical technologies aiming at Earth observation, tests of fundamental physics and gravitational wave detection. Proposed and projected space missions like STE-QUEST \cite{Aguilera2014}, CAL \cite{Dincao2017,CAL}, QWEP \cite{TINO2013}, Q-TEST \cite{Williams2016} aim at the observation of Bose-Einstein condensates at unprecedented expansion times, quantum gas physics in the pico Kelvin regime, and dual-species atom interferometry for future precision tests of the equivalence principle with quantum matter \cite{Stamminger2015,Aguilera2014,Dincao2017,CAL}. Such experiments involving light-atom interaction, e.g., for laser cooling or atom interferometry, require laser systems whose optical frequency is stabilized to specific atomic transitions. Moreover, precise frequency control with high demands on frequency stability and agility as well as intensity control is mandatory.
Planned gravitational wave observatories, such as LISA \cite{Vitale2014}, use inter-satellite laser ranging with laser systems at \SI{1064}{\nano\meter} for detection of gravitational waves in a spectral window between \SI{0.1}{\milli\hertz} and \SI{1}{\hertz}. Next generation gravity missions (NGGM) might use similar laser ranging techniques for global mapping of temporal variations of Earth's gravitational field \cite{Kawamura2008a}. 
The requirement on laser frequency noise of these missions can be achieved by laser frequency stabilization to optical cavities or atomic or molecular transitions. Related technologies have been or are currently being developed in the context of missions aiming at space-borne atom interferometry \cite{Schuldt:17,Schkolnik2016,Dinkelaker:16,Lezius:16,Leve2015,Leveque2014} or atomic clocks \cite{Aguilera2014,Laurent2015540,Swierad2016}.

In addition to the qualification in environmental testing facilities, the deployment of laser systems in realistic scenarios and relevant environments, offered by sounding rockets or zero-g parabolic flights, allows for rapid iterative tests and further development of related technologies. Sounding rocket mission in particular close the gap between ground and space applications \cite{Schkolnik2016,Dinkelaker:16} but also enable scientific, pioneering pathfinder experiments as shown with recent MAIUS mission, demonstrating the first realization of a Bose-Einstein condensate of $^{87}$Rb in space.
Sounding rocket missions based on VSB 30 motors, that are used in the TEXUS program, allow for a \SI{16}{\minute} ballistic flight reaching an apogee of about \SI{250}{\kilo\meter} after \SI{4.2}{\minute} ascent, followed by about \SI{6}{\minute} of $\mu g$ time \cite{Grosse2014}. Typically, four experiment modules can be launched altogether on one mission as independent payloads, each with a diameter of \SI{43.8}{cm} that can be covered in a pressurized dome. The total scientific payload mass is usually limited to \SI{260}{kg}, with a total payload length of \SI{3.4}{m}.

In the JOKARUS mission, we aim to demonstrate an absolute optical frequency reference at \SI{1064}{\nano\meter} on a sounding rocket.
The JOKARUS laser system is based on modulation transfer spectroscopy of the hyperfine transition R(56)32-0:a$_{10}$ in molecular iodine at \SI{532}{\nano\meter}, using a frequency-doubled extended cavity diode laser (ECDL). Iodine frequency standards realized with frequency doubled Nd:YAG lasers locked to this hyperfine transition have been investigated in detail for many years as optical frequency standards \cite{Ye1999,Nevsky2001a}.
Thanks to the strong absorption and narrow natural linewidth of \SI{144}{\kilo\hertz} \cite{Eickhoff1995a} these systems exhibit fractional
frequency instabilities as low as \num{3e-15}\cite{Schuldt:17} and an absolute frequency reproducibility of few kHz \cite{Hong2001b}. 
These features make them promising candidates for future space missions targeting at the detection of gravitational waves, such as LISA, or monitoring of Earth's gravitational potential, such as NGGM \cite{Kawamura2008a}, which rely on laser-interferometric ranging with frequency-stable laser systems at \SI{1064}{\nano\meter} distributed on remote satellites that need to be precisely synchronized.
Different realizations of iodine references for space missions were proposed and investigated on a breadboard level \cite{Musha2012,Leonhardt2006} and prototypes were built that fulfill the  requirements on the frequency stability for such missions \cite{Argence2010,Schuldt:17,Schuldt:16a}. 
The JOKARUS mission will, for the first time, demonstrate an autonomous, compact, ruggedized iodine frequency reference using a micro-integrated high power ECDL during a space flight.  

\section{Status of laser system qualification on sounding rockets}
\label{sec:heritage}

\begin{figure}[tbhp]
\centering
\includegraphics[width=0.99\linewidth]{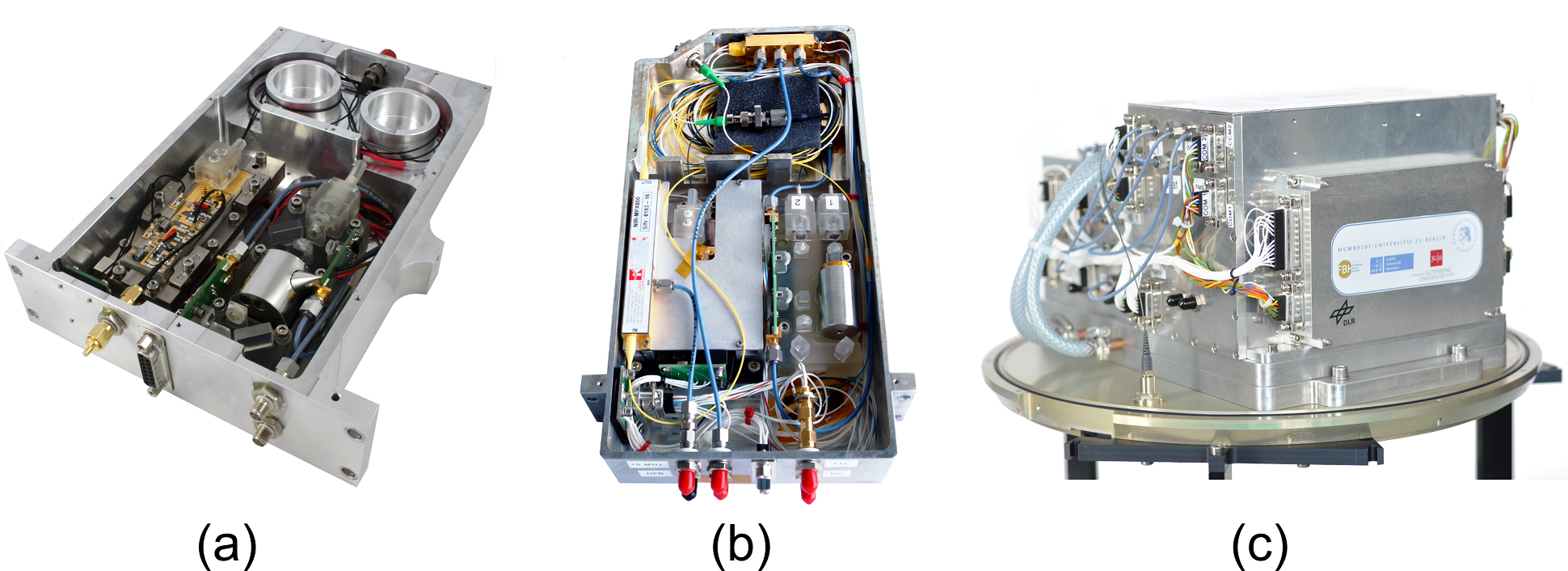}
\caption{Photographs of three laser payloads successfully operated on sounding rocket flights.(a) FMS based DFB laser system for Rb spectroscopy (FOKUS, \SI{215}{mm}$\times$\SI{170}{mm}$\times$\SI{50}{mm}, \SI{2.1}{kg}), (b) MTS based DBF reference laser system for Rb spectroscopy (FOKUS-Reflight, \SI{288}{mm}$\times$\SI{170}{mm}$\times$\SI{50}{mm}, \SI{3.0}{kg}) and (c)
Potassium stabilized laser system with two ECDLs (KALEXUS, \SI{345}{mm}$\times$\SI{218}{mm}$\times$\SI{186}{mm}, \SI{16}{kg}).}
\label{fig:heritage}
\end{figure}

In three successful rocket missions, namely FOKUS \cite{Lezius:16}, FOKUS Re-Flight and KALEXUS \cite{Dinkelaker:16}, we and partners have demonstrated the maturity of our laser systems and related technologies \cite{Duncker:14}. As part of the FOKUS mission,  flown on the 23\textsuperscript{rd} of April 2015, a frequency-stabilized laser system, shown in Fig.\,\ref{fig:heritage}(a), was qualified as master laser for the MAIUS laser system \cite{Schkolnik2016}.
It is based on frequency modulation spectroscopy (FMS) of the $\mathrm{D}_2$ transition in $^{87}$Rb using a micro-integrated distributed feedback (DFB) laser module \cite{Schiemangk2015}.
FOKUS demonstrated the first optical Doppler-free spectroscopy in space. Moreover, the laser frequency was compared to a Cesium (Cs) reference using an optical frequency comb (\textit{Menlo Systems}) during flight, making the FOKUS mission a demonstrator for a null test of the gravitational red shift between the optical transition in Rb and a microwave Cs clock \cite{Lezius:16}. The FOKUS payload was flown again on the 17\textsuperscript{th} of January 2016 aboard the TEXUS 53 sounding rocket under the name FOKUS Re-Flight together with KALEXUS. FOKUS Re-Flight, shown in Fig.\,\ref{fig:heritage}(b), was updated from FOKUS to a system based on modulation transfer spectroscopy (MTS) using a fiber-pigtailed phase modulator. The KALEXUS mission featured two micro-integrated extended cavity diode lasers (ECDL) \cite{Luvsandamdin:14} operating at \SI{767}{nm} that were alternately offset-locked to each other and stabilized to potassium using FMS. This way, redundancy and autonomy concepts for future space missions were demonstrated \cite{Dinkelaker:16}. The successful FOKUS and KALEXUS missions constitute important qualification steps for the MAIUS mission and towards its follow on missions MAIUS II and III aiming at dual species atom interferometry with Rb and K in space as well as planned satellite missions. 

JOKARUS is based on the laser system heritage of these successful sounding rocket missions and is planned to operate on a sounding rocket mission end of 2017. The next section describes the JOKARUS system design and the estimated performance in context of laser interferometric ranging missions.

\section{JOKARUS Payload---System Design}
\label{sec:payload}

\begin{figure}[tbh]
\centering
\includegraphics[width=0.99\linewidth]{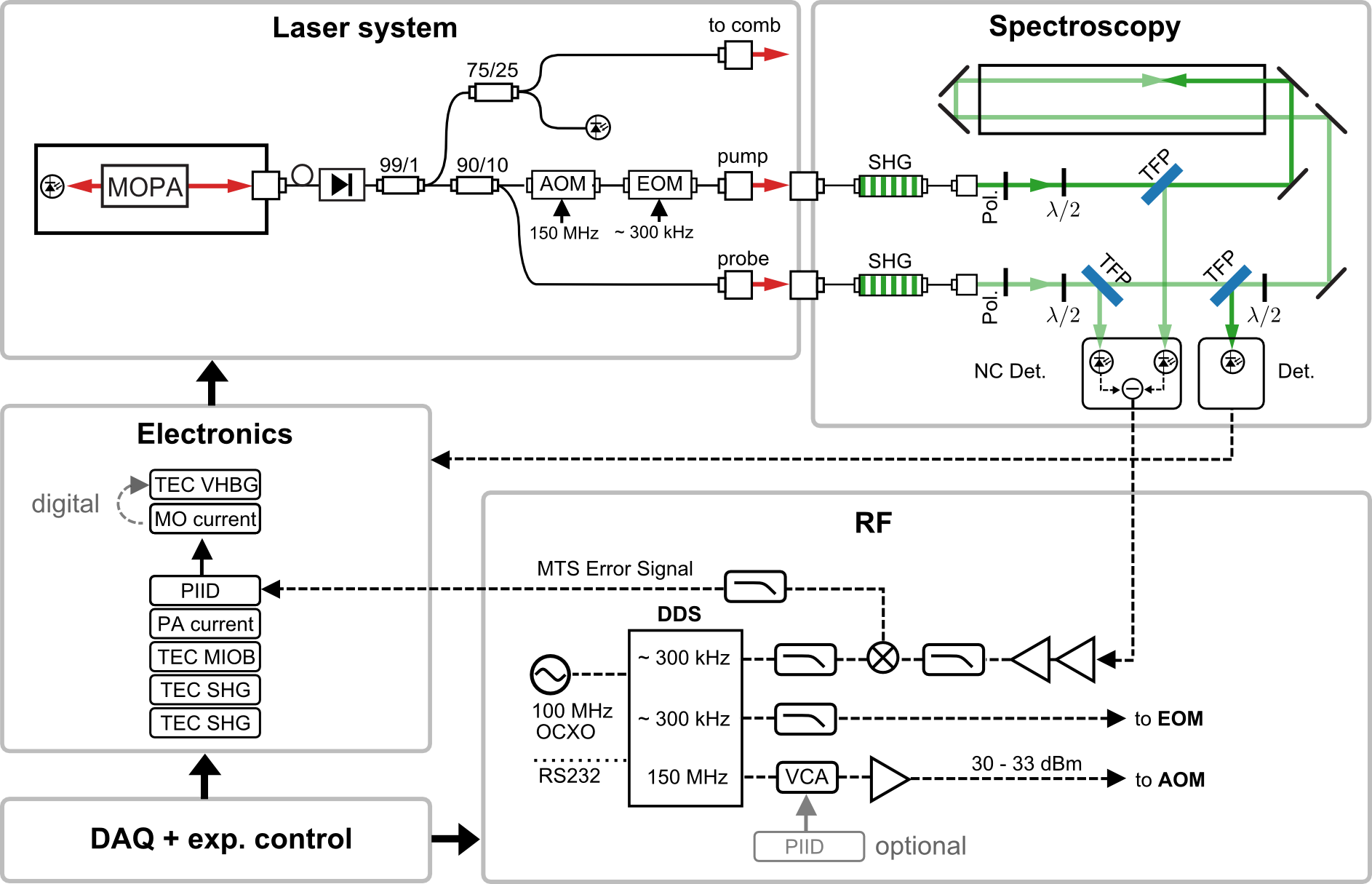}
\caption{Scheme of the JOKARUS payload including the laser system, spectroscopy and RF, electronics and experiment control. See text for details.}
\label{fig:JOKARUS}
\end{figure}

The JOKARUS payload, schematically shown in Fig.\,\ref{fig:JOKARUS}, features a laser system based on a ECDL with an integrated optical amplifier operating at \SI{1064}{nm}, a spectroscopy module including the quasi-monolithic setup for modulation transfer spectroscopy of molecular iodine and two periodically poled lithium niobate (PPLN) waveguide modules for second harmonic generation (SHG), as well as RF and control electronics for frequency stabilizing the ECDL laser to the spectroscopy setup. 
The individual subsystems are presented in the following sections.

\subsection{Laser System}
\label{subsec:laser}

\begin{figure}[b]
\centering
\includegraphics[width=0.99\linewidth]{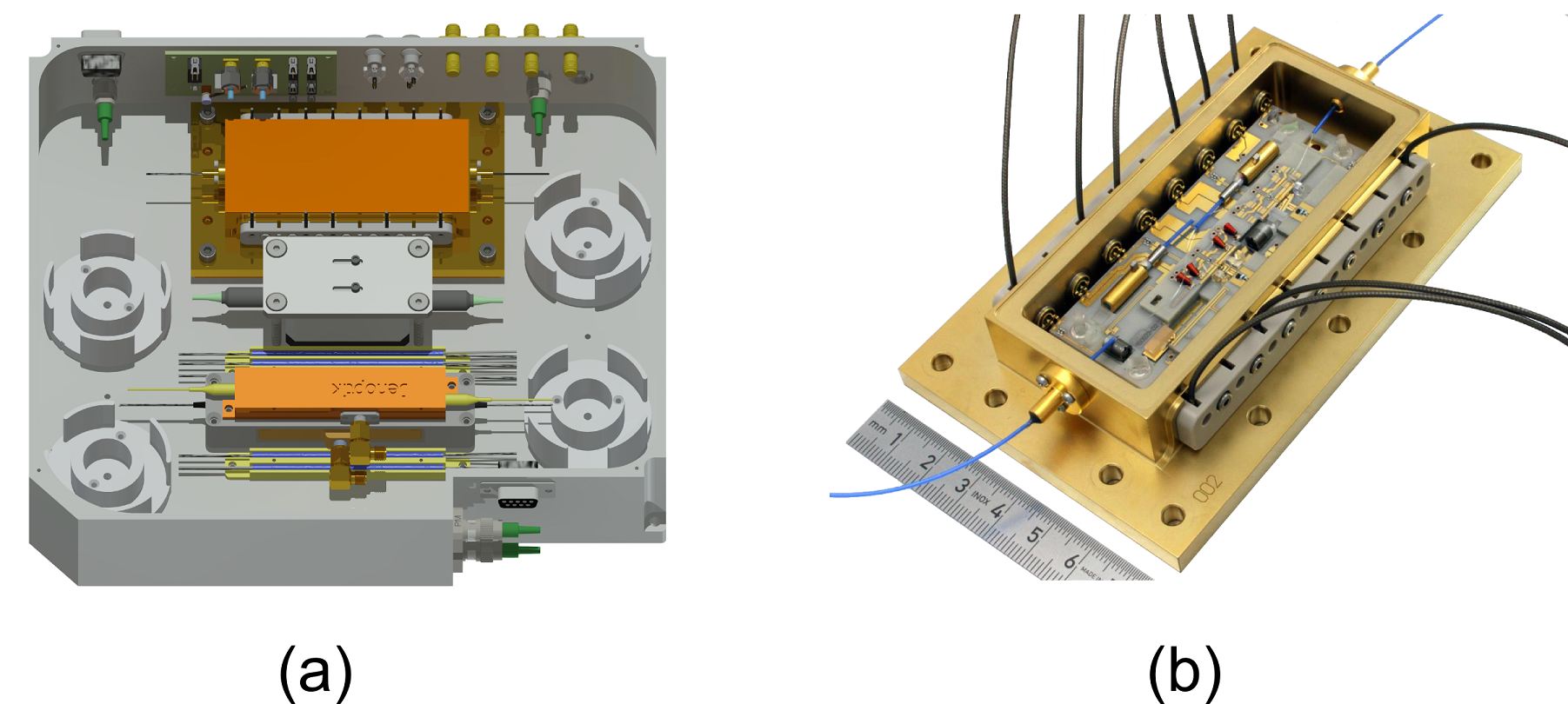}
\caption{(a) CAD drawing of the JOKARUS laser system module (Laser system in Fig.\,\ref{fig:JOKARUS}). From top to bottom: ECDL, optical isolator, EOM and AOM.  The dimensions are \SI{290}{mm} $\times$ \SI{225}{mm} $\times$ \SI{58}{mm} with a total mass of \SI{3.5}{kg}. (b) A photograph of the assembled JOKARUS ECDL with dimensions of \SI{128}{mm} $\times$ \SI{78}{mm} $\times$ \SI{23}{mm} and a total mass of \SI{760}{g}.
}
\label{fig:laserebene}
\end{figure}

The laser system is housed in a module as shown in Fig.\,\ref{fig:laserebene} a) and includes the laser, an electro-optic modulator (EOM) and an acousto-optic modulator (AOM) for preparation of the spectroscopy beams. The laser is a micro-integrated master-oscillator-power-amplifier-module (MOPA) developed and assembled by the Ferdinand-Braun-Institut (FBH), see Fig.\,\ref{fig:laserebene} b). The MOPA consists of a narrow-linewidth ECDL master oscillator operating at \SI{1064}{nm} and a high-power power amplifier.
A previous generation of the laser module is described in \cite{Luvsandamdin:14} and details on the performance of the laser module and the technology applied for its integration will be given elsewhere \cite{Kuerbis2017}.
The ECDL master oscillator provides an emission linewidth of less than \SI{50}{kHz} (\SI{1}{ms}, FWHM) and allows for high bandwidth frequency control via control of the injection current. 
The laser module provides a fiber-coupled optical power of \SI{500}{\milli\watt} at an injection current of \SI{1500}{\milli\ampere}. This module was subject to a vibration qualification according to the requirements of the TEXUS program (\SI{8.8}{g_{rms}}), while similar laser modules of previous generations even passed \SI{29}{g_{rms}} and \SI{1500}{g} pyroshock tests \cite{Luvsandamdin:14}.

In the laser system, see Fig.\,\ref{fig:JOKARUS}, the MOPA is followed by an optical isolator (\textit{Thorlabs}) and a 99:1 fiber splitter, where few mW are separated from the main beam for frequency measurements with the frequency comb that is part of another payload aboard the TEXUS\,54 mission. A second fiber splitter divides the main beam into pump and probe beam for modulation transfer spectroscopy (MTS). The probe beam is connected to a fiber-coupled PPLN wave guide module (\textit{NTT Electronics}) for second harmonic generation.
The pump beam is frequency shifted by \SI{150}{\mega\hertz} using a fiber-coupled AOM (\textit{Gooch \& Housego}) to shift spurious interference between pump and probe beam outside the detection bandwidth of the MTS signal.
The frequency-shifted beam is then phase-modulated at $\approx$ \SI{300}{\kilo\hertz} by a fiber-coupled EOM (\textit{Jenoptik}) and will later also be frequency-doubled by a second SHG module.
Taking nominal losses of the components, splice connections and the conversion efficiency of the SHG modules into account, we expect an optical power of \SI{10}{\milli\watt} and \SI{3}{\milli\watt} for the pump and probe beam at \SI{532}{\nano\meter}, respectively, which is sufficient for saturation spectroscopy . The power of the pump beam can be stabilized by using a voltage controlled attenuator (VCA) and a feedback loop (see Fig.\,\ref{fig:JOKARUS}). 
Several fiber taps are used for power monitoring at various positions in the laser system during flight. The pump and probe beam are finally guided from the laser module to the spectroscopy module by polarization maintaining fibers at \SI{1064}{\nano\meter} and mating sleeves connectors.

All components of the laser system were qualified at a random vibration level of \SI{8.8}{g_{rms}} (hard-mounted) to ensure their integrity after the boost phase of the rocket launch.

\subsection{Spectroscopy Module}
\label{subsec:spec}

\begin{figure}[tbhp]
\centering
\includegraphics[width=0.99\linewidth]{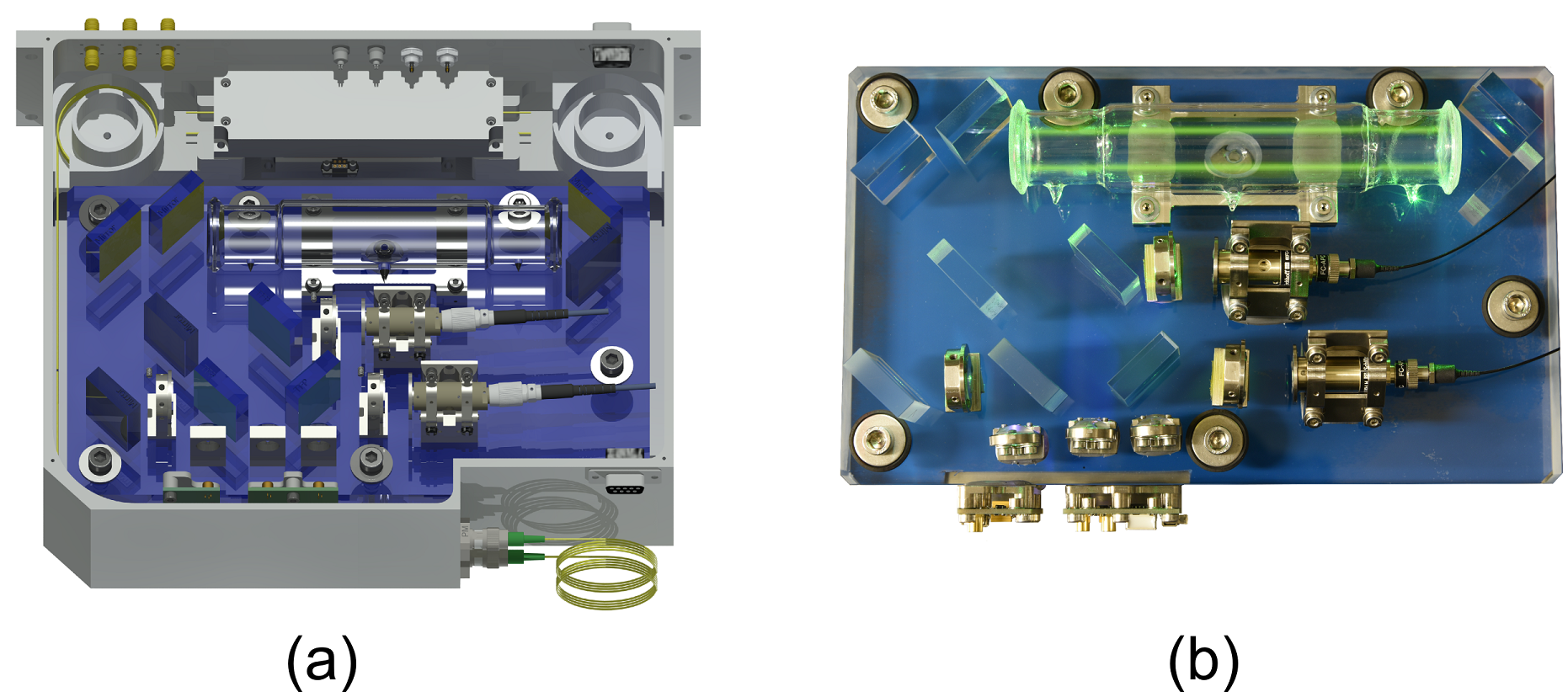}
\caption{(a) CAD drawing of the JOKARUS spectroscopy module (Spectroscopy in Fig.\,\ref{fig:JOKARUS}). The two SHG modules are located at the top. The dimensions are \SI{290}{mm} $\times$ \SI{225}{mm} $\times$ \SI{78}{mm} with a total mass of \SI{6}{kg}. (b) Photograph of the assembled JOKARUS spectroscopy setup.}
\label{fig:Spectroscopy}
\end{figure}

The spectroscopy setup is housed in a separate module as shown in Fig.\,\ref{fig:Spectroscopy} together with the SHG modules (cf. Fig.\,\ref{fig:JOKARUS}). It is based on previous iterations of an iodine reference for deployment in space missions, developed at ZARM Bremen, DLR Bremen and the Humboldt-Universit\"at zu Berlin \cite{Schuldt:17}.
The optical setup is realized using a special assembly integration technique \cite{res10}, where the optical components are bonded directly on a base plate made from fused silica with a footprint of \SI{246}{\mm} $\times$ \SI{145}{\milli\meter} resulting in a quasi-monolithic, mechanically and thermally stable spectroscopy setup as shown in Fig.\,\ref{fig:Spectroscopy} (b).
An iodine setup using this assembly technique was subjected to environmental tests including vibrational loads up to 29\,g$_\text{rms}$ and thermal cycling from \SI{-20}{\celsius} to \SI{+60}{\celsius} \cite{Schuldt:16a}. 

In the JOKARUS MTS setup, the pump beam is launched from a fiber collimator (\textit{Sch\"after}\,+\,\textit{Kirchhoff}) with a beam diameter of \SI{2}{\milli\meter} and is guided twice through an iodine cell with a length of \SI{15}{\cm}, resulting in an absorption length of \SI{30}{\centi\meter}.
Behind the cell, the pump beam is reflected at a thin film polarizer (TFP) and focused on a photo detector for optional power stabilization.
The probe beam is launched with the same beam diameter and is split using a TFP into a probe and a reference beam for balanced detection using a noise-canceling detector adapted from \cite{Hobbs1997a}, as shown in Fig.\,\ref{fig:JOKARUS}. 
The iodine cell is provided by the Institute of Scientific Instruments of the Academy of Sciences of the Czech Republic (ISI) in Brno, filled with an unsaturated vapor pressure of $\approx$\SI{1}{\pascal} \cite{Hrabina2014b}.

\subsection{Electronics}
\label{subsec:electronics}

The electronic system for JOKARUS is segmented in 3 functional units. First, the RF electronics for the optic modulators shown in Fig.\,\ref{fig:JOKARUS}. It is based on a direct digital synthesizer (DDS9m, \textit{Novatech Instruments}), referenced to an oven-controlled crystal oscillator.
The DDS provides a \SI{150}{\mega\hertz} signal for the AOM and two signals for phase modulation of the pump beam via the EOM and analog demodulation of the MTS signal.
Second, a stack of compact electronic cards by \textit{Menlo Systems} based on the FOKUS flight electronics \cite{Lezius:16} are used for temperature control of the SHG modules and the diode laser, as current source for the ECDL-MOPA and  for realizing the feedback control for laser frequency stabilization. The cards are controlled  by an ARM based embedded system via a CAN interface, also providing an interface to higher level data acquisition.
The third unit contains a 16--bit DAQ card that is used for data acquisition and the x86-based flight computer (\textit{exone IT}). It runs the experiment control software that provides coarse tuning of the laser frequency, identification of the fine transition R(56)32-0 as well as invoking and controlling a PID feedback control for frequency stabilization to the selected hyperfine transition. 
 
\subsection{Payload Assembly}

\begin{figure}[tbhp]
\centering
\includegraphics[width=0.99\linewidth]{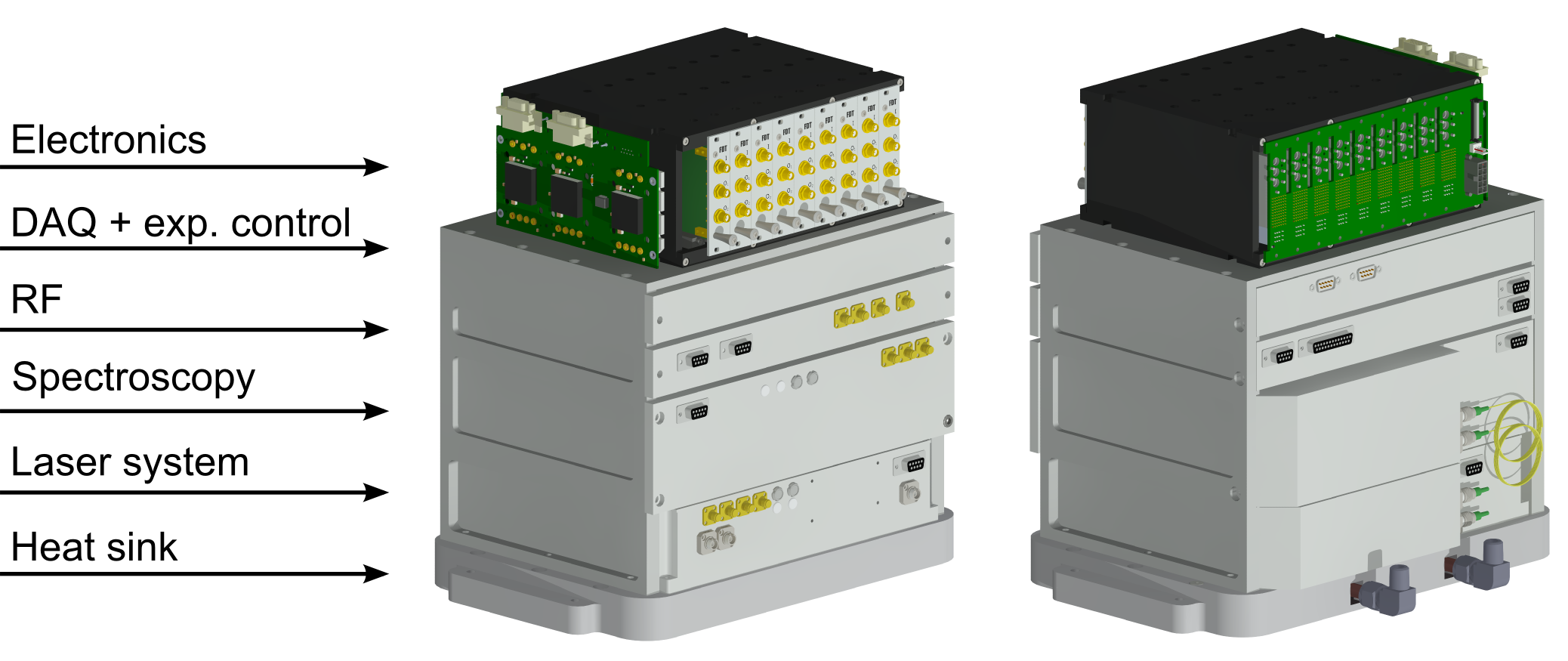}
\caption{CAD drawing of the JOKARUS payload. The payload dimensions are \SI{345}{mm} $\times$ \SI{270}{mm} $\times$ \SI{350}{mm} with a total mass of \SI{25}{\kilo\gram} and a power consumption of \SI{100}{W}. 
}
\label{fig:JOKARUS_CAD}
\end{figure}

The subsystems presented above are integrated in individual housings made from aluminum that share a common frame as a support structure shown in Fig.\,\ref{fig:JOKARUS_CAD}. A water-cooled heat sink is integrated into the base frame for temperature control until liftoff. During flight, we expect an average temperature increase of about \SI{3}{K} throughout the mechanical structure, based on nominal power consumption of \SI{100}{W}. The optical fiber connection between the laser and spectroscopy units are realized via mating sleeves. The total payload has a dimension of \SI{345}{mm} $\times$ \SI{270}{mm} $\times$ \SI{350}{mm} and a total mass of \SI{25}{\kilo\gram}, which allows for integration into the TEXUS sounding rocket format.

\begin{figure}[tbhp]
\centering
\includegraphics[width=0.80\linewidth]{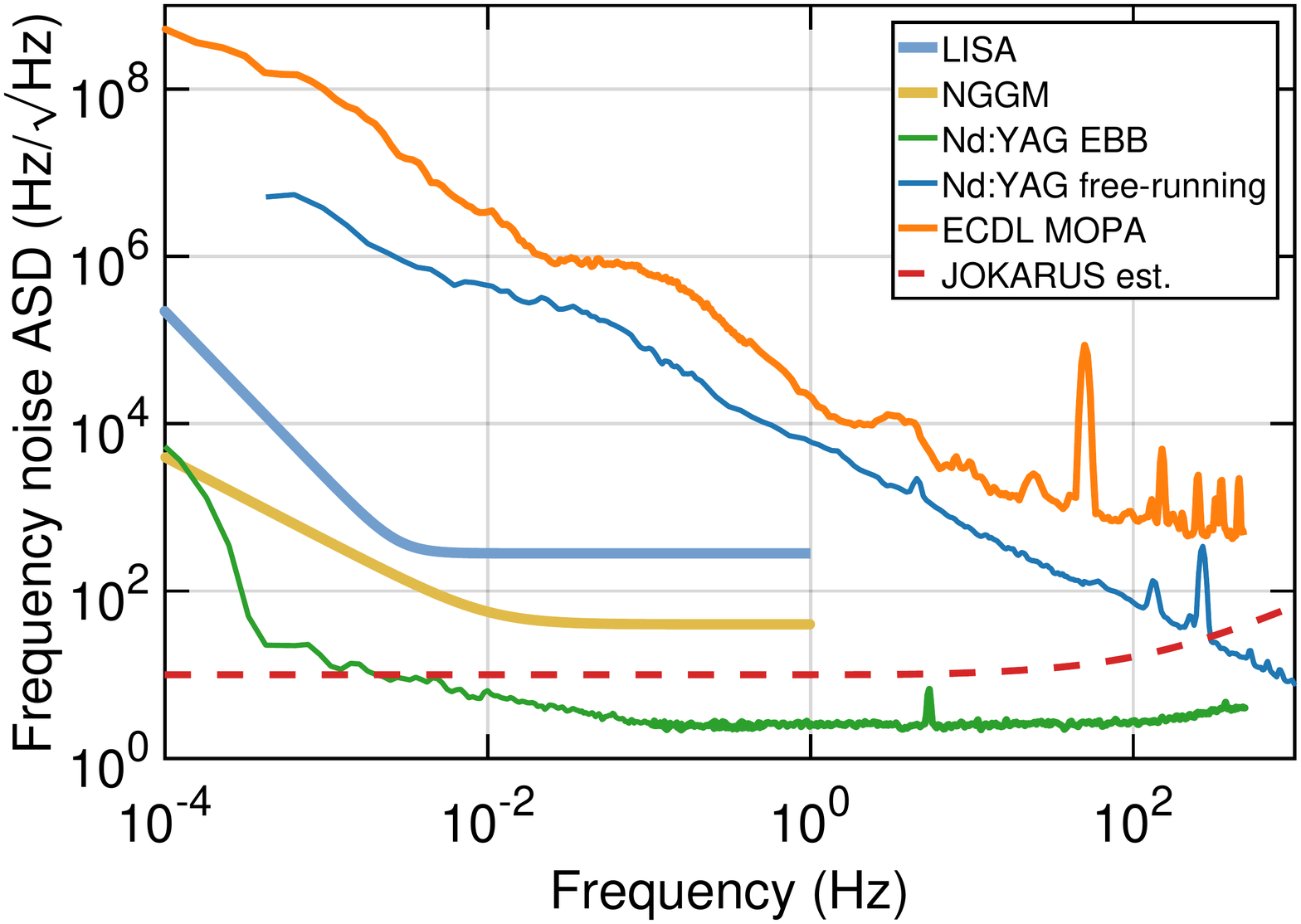}
\caption{Estimated frequency stability of JOKARUS in comparison to another iodine frequency reference (EBB) using an Nd:YAG laser. Frequency noise of the free-running Nd:YAG laser (blue) is suppressed by frequency stabilization to the EBB resulting in frequency noise depicted in green. The frequency noise of the ECDL is expected to be reduced to the frequency noise indicated by the red dashed line when locked to the JOKARUS spectroscopy module. The requirements on the laser frequency noise for missions like LISA and NGGM are shown in blue and yellow, respectively.}
\label{fig:JOKARUS_estitmation}
\end{figure}

We estimated the performance of the JOKARUS system in terms of amplitude spectral density (ASD) of frequency noise in comparison to the frequency stability of an iodine reference developed, characterized and reported previously \cite{Schuldt:17,Schuldt:16a}, called elegant breadboard model (EBB).
The frequency noise achieved with the EBB is shown in Fig.\,\ref{fig:JOKARUS_estitmation} (green graph) together with the frequency noise of the free-running Nd:YAG laser (blue graph) used in this setup.
The EBB fulfills the requirement on the frequency noise of planned space missions like LISA and NGGM. 
For the JOKARUS instrument we expect to achieve a frequency noise on a level of \SI[per-mode = symbol]{10}{\hertz\per\hertz\tothe{1/2}} (red dashed graph) which corresponds to a fractional frequency instability of  \num{2.4e-14}$/\sqrt{\tau}$. The performance was estimated from the frequency noise of the free-running ECDL (orange graph) using a control bandwidth of $\approx$\SI{100}{\kilo\hertz} and taking into account a factor of three shorter absorption length in JOKARUS compared to the EBB.
We therefore expect the JOKARUS instrument to fulfill the requirements on the frequency noise of space missions like LISA and NGGM.

\section{Conclusion}
\label{sec:summary}

We presented the design of an absolute optical frequency reference developed as a payload for a sounding rocket mission and a potential candidate for space applications that require frequency-stable laser systems at \SI{1064}{\nano\meter}. The payload will be part of a sounding rocket mission planned for a launch end of 2017, where the optical frequency of the JOKARUS frequency reference will be compared to an optical frequency comb during a \SI{6}{\minute} space flight. JOKARUS will demonstrate autonomous operation of an absolute optical frequency reference whose performance is expected to meet the requirement on the frequency noise of laser systems for future space missions such as LISA or NGGM. 

The assembly integration technology used for the iodine spectroscopy setup is a very promising technology for realization of compact and ruggedized space optical systems. Future space missions such as NGGM, LISA or MAQRO \cite{Kaltenbaek2016} can benefit from this technology heritage.

As demonstrated with the FOKUS and KALEXUS missions operating laser frequency references at \SI{780}{\nano\meter} and \SI{767}{\nano\meter}, respectively, this project further shows the versatility of the micro-integrated diode laser technology for the realization of compact and efficient laser systems for applications in the field.   
Future space missions using laser or atom interferometry or the development of space optical clocks may benefit from this technology heritage and its first applications in space missions.

\section*{Authors contributions}

VS conceived and designed the payload and authored this manuscript.
KD conceived the conception and design of the laser system and spectroscopy setup, estimated system performance and authored this manuscript. FBG conceived the experiment control concept. MO, TS and CB conceived and designed the iodine spectroscopy setup. ML and RH conceived and designed the electronics. AB and CK contributed to the concept of the laser module. MK and AP conceived the payload.\\
All authors revised this manuscript.

\section*{Acknowledgments}

The authors thank Ulrich Johann and Alexander Sell from Airbus DS (Friedrichshafen) for support within the Laboratory of Enabling Technologies where the iodine spectroscopy unit was integrated.
This work is supported by the German Space Agency DLR with funds provided by the Federal Ministry for Economic Affairs and Energy under grant numbers  DLR\,50WM\,1646, 50\,WM\,1141, 50\,WM\,1545.

\end{document}